\documentclass[twoside]{article}
\usepackage{fleqn,espcrc2}

\newcommand{\AmS}{{\protect\the\textfont2
  A\kern-.1667em\lower.5ex\hbox{M}\kern-.125emS}}

\title{Dirac Branes, Characteristic Currents and Anomaly Cancellations in 
5-Branes}

\author{K. Lechner\address{Department of Physics and INFN Sez.
Padova
- Via F.Marzolo,8 - 35131 Padova, Italy} and %
P. A. Marchetti\addressmark\thanks{This work was supported by
the European Commission RTN 
        programme HPRN-CT2000-00131. Talk presented by P.A. Marchetti.}
        }
        
\begin{document}

\begin{abstract}
The aim of this note is to discuss, in a  somewhat informal language, 
the cancellation of  anomalies (in topologically trivial space-time) for 
5-branes using as "building blocks":
i) a generalization to $p$-branes of the Dirac strings of monopoles (Dirac 
branes) and a refinement of this idea involving a geometric regularization of 
Dirac branes, leading to the formalism of "characteristic currents"
ii) the PST formalism .
As an example of the potentiality of the developed framework we discuss in 
some detail the anomaly cancellation in the D=10 effective theory of 
heterotic string and 5-brane coupled to supergravity, where the anomaly 
inflow is automatically generated. Some remarks are also made on a similar 
approach to the problem of anomaly cancellation in the effective theory 
of M5-brane coupled to D=11 supergravity, developed in collaboration with 
M.Tonin, where however still as open problem remains
a Dirac anomaly.

\vspace{1pc}
\end{abstract}

\maketitle

\section{Currents}

P-branes, Dirac strings and their generalizations found a natural 
mathematical formulation in terms of (de Rham) currents: a $p$-current in 
a $d$-dimensional space $M$ is a "$p$-form whose coefficients are 
distributions" or, more precisely, a linear functional on the space of 
smooth ($d-p$)-forms with compact support, continuous in the sense of 
distributions \cite{Rham}.
As an example let us consider $M= {\bf R}^{4}, \gamma$ a curve, which one 
might identify as the worldline of a charged particle, parametrized in terms 
of the  proper time $\tau$. There is a  3-current naturally associated 
to $\gamma$:

\begin{eqnarray}\label{1}
j(x) &=& j_{\mu\nu\rho}  (x) dx^\mu \wedge dx^\nu \wedge dx^\rho \\
j_{\mu\nu\rho}(x) &=& \epsilon_{\mu\nu\rho\sigma} 
\int {dx^\sigma (\tau)\over d\tau}
\delta^{(4)} (x-x(\tau)) d\tau.\nonumber
\end{eqnarray}

This example also justifies the term "current" because, with the physical 
identification alluded above, $j$ is really the (Hodge dual of the) 
electromagnetic current generated by a charged particle of unit charge.
Obviously, in particular, all smooth $p$-forms are $p$-currents.

Moving from to $p$-forms to $p$-currents we gain a very useful extension of 
Poincar\'e duality.
We recall that Poincar\'e duality for forms \cite{bott} is a map from the 
homology class of a $k$-cycle ("closed $k$-surface") $c$ into the cohomology 
class of a $d-k$ closed form $\alpha_c$ such that for any closed $k$-form 
$\beta$ of compact support 

\begin{equation}\label{2}
\int_c \beta=\int_M \alpha_c \wedge \beta.
\end{equation}

In the space of currents there is a natural map from $k$-chains 
("$k$-surfaces") $\Sigma_k$ to $d-k$ currents, $PD$ $(\Sigma_k)$, 
generalizing Poincar\'e duality in the sense that 

\begin{equation}\label{3}
\int_{\Sigma_k} \beta= \int_M PD (\Sigma_k) \wedge \beta,
\end{equation}
for every smooth $k$-form $\beta$ of compact support.

It is clear that, loosely speaking, the coefficients of $PD (\Sigma_k)$ are 
$\delta$-functions with support on $\Sigma_k$ as in the previous example 
where $j=PD (\gamma)$.
In particular, the image by $PD$ of a closed surface is a closed current and 
\begin{equation}\label{4}
PD (\partial\Sigma_k) = d PD (\Sigma_k), 
\end{equation}
where $\partial$ denotes the boundary operator.  
$k$-currents which are linear combinations with integer coefficients of 
$PD$ of $k$-chains are called integral currents.
If well defined, an integral like

\begin{equation}\label{5}
\int PD(\Sigma_k) \wedge PD (\Sigma_{d-k})
\end{equation}
%\subsection{Spacing}
is an integer, the intersection number of $\Sigma_k$ and $\Sigma_{d-k}$.

\section{Dirac branes}
Let us show how one uses currents to describe Dirac strings or their 
generalizations. 
Suppose one has an equation of motion for an invariant curvature $H$, e.g. 
the field strength of a monopole,

\begin{equation}\label{6} 
dH =j,
\end{equation}
where $j=PD (\Sigma)$ for some worldvolume $\Sigma$ of a closed brane.
A generalization of Poincar\'e lemma \cite{Rham} then proves that there 
exists a 
surface $S$ such that $\Sigma= \partial S$, so that by (\ref{4}) 

\begin{equation}\label{7}
j=dC , \quad C= PD (S).
\end{equation}

The surface $S$ is defined modulo a boundary, i.e. we can replace $S$ by 
$S' = S+\partial V$, then accordingly we replace $C$ by 
\begin{equation}\label{8}
C' = PD (S') = C+ PD(\partial V) = C+d PD (V).
\end{equation}

It is clear that if $H$ is the field strength of a monopole and $j$ the 
generated magnetic current in $d=3+1$, then $C$ may be identified as the 
worldvolume of its Dirac string. Hence in general if $j$ is a 
$(d-p)$-current describing the worldvolume of a $(p-1)$-brane, we say that 
the corresponding $(d-p-1)$-current $C$ describes the worldvolume of a 
$p$-Dirac brane. 
In terms of this $p$-Dirac brane, one can solve (\ref{6}), in a topologically 
trivial space-time, by
\begin{equation}\label{9}
H=dA+C,
\end{equation}
where $A$ is a $d-p-2$ gauge form, and  must be interpreted in a 
distributional sense, i.e. as a $d-p-2$-current. The  "curvature" (\ref{2}) 
is invariant under the change of
Dirac brane:

\begin{equation}\label{10}
A \rightarrow A - PD (V), \ \ \ C \rightarrow C + d PD (V),
\end{equation}
where $V$ is a $p+2$ surface. 

\section{Characteristic currents}

Actually we need a refinement of the notion of current because in 
supergravity (SUGRA) one has to consider what naively can be interpreted 
as the restriction of an integer current to its support, which is clearly  
mathematically ill defined. As an example consider, in $d=4$, a surface 
current  
$j= PD (\Sigma)$ where $\Sigma$ is a 2-surface parametrized by 
$\tau \equiv \{\tau_1, \tau_2\}$, which admits a local representation 
analogous to (\ref{1}):
\begin{eqnarray}\nonumber 
j (x) &=& j_{\mu\nu} (x) dx^\mu \wedge  dx^\nu \\
j_{\mu\nu} (x) &=& \epsilon_{\mu\nu\rho\sigma} \int {dx^\rho (\tau)\over 
d\tau_1} {dx^\sigma (\tau)\over d\tau_2} \cdot \\
&\delta^{(4)}& (x-x(\tau)) d^2\tau \nonumber
\end{eqnarray}  
Then in 
$j (x) |_\Sigma$ the term $\delta^{(4)} (x-x(\tau)) |_\Sigma \sim \infty$ for 
$x\in \Sigma$ and the term $dx^\mu \wedge dx^\nu \wedge 
dx^\rho |_\Sigma \sim 0$.
To deal with this problem we need some preliminary remarks. We recall that 
a tubular neighbourhood $N(X)$ of a closed submanifold $X$ of codimension $k$ 
is isomorphic to the normal bundle $N$ of $X$ which is an $SO(k)$-bundle 
over $X$. 
Mathematicians call characteristic current \cite{Harv} in an $SO(k)$-bundle 
over $X$ a family of smooth closed currents $j^\epsilon, \epsilon > 0$ on 
the total space of the bundle, regularizing $PD(X)$,  satisfying:
\begin{equation}\label{11}
j^\epsilon |_X = \chi_k (N), \ \ \lim_{\epsilon\rightarrow 0}j^\epsilon = 
PD (X)
\end{equation}
where $\chi_k(N)$ denotes the $k$-Euler form of the $SO(k)$-bundle.
We need a slightly weaker definition where the first equation in (\ref{11}) 
is replaced by 
\begin{equation}\label{12}
\lim_{\epsilon\rightarrow 0} j^\epsilon|_X = \chi_k (N). 
\end{equation}
By abuse of language we still refer to such $j^\epsilon$ as a 
characteristic  current. One can easily verify that in a flat space-time a 
characteristic current is obtained by replacing the $\delta$-function 
appearing in the definition of $j = PD(X)$ by a gaussian:
\begin{equation}\label{13}
\delta (x - x (\tau) ) \rightarrow {e^-{(x-x(\tau))^2 \over \epsilon}\over 
(\pi\epsilon)^{d/2}}
\end{equation}
and identifying the bundle $N$ with $N(X)$.
For characteristic currents we have
\begin{equation}\label{14}
j^\epsilon \wedge j^\epsilon \rightarrow_{\epsilon\rightarrow 0}\chi_k (N)
\wedge PD (X).
\end{equation}

We remark that by means of the regularization (\ref{13}) 
one can extend the notion of characteristic current also to non-closed 
submanifolds and via  covariantization to  non flat spaces-time. 
Furthermore since this regularization corresponds to a convolution, 
it commutes with the exterior differential $d$.

\section{PST formalism}
The last ingredient we need is the Pasti-Sorokin-Tonin (PST) 
formalism \cite{Past} to deal with duality conditions, e.g. for invariant 
curvatures of the rank 2 and 6 Ramond-Ramond fields in  type IIA SUGRA 10:

\begin{equation}\label{15}
H_3 = * H_7,
\end{equation}
or self-duality conditions, e.g. for the invariant curvature of the 2-form 
on the worldvolume of the M 5-brane in SUGRA 11: 
\begin{equation}\label{16} 
h_3 = * h_3.
\end{equation}
The PST approach allows to write, in a topologically trivial space time, 
an action which is local, although non-polynomial, diffeomorphism invariant 
and which permits to derive duality or self-duality conditions as equations 
of motion. Let us sketch the basic idea, for details see \cite{Past}.
(To simplify the notation from now on the wedge product symbol $\wedge$ is 
understood and some sign depending on the dimension $d$ and the rank of the 
form is omitted, see \cite{Lech}).

One introduces a 0-form $a$ and defines the 1-form 
\begin{equation}\label{17}
v (x) = {da (x)\over ||da||(x)}, 
\end{equation}
where $||da||^2 = *(da * da)$.
Let us assume that we are dealing with a  model involving a pair of forms 
with invariant curvatures of rank $k$ and $d-k$, denoted by $F^1$ and $F^2$, 
satisfying a duality condition
\begin{equation}\label{16a}
F^1 = * F^2, \quad F^\alpha = dA^\alpha + C^\alpha, 
\end{equation}
and whose dynamics is determined by a "Maxwell" like action.
Defining
\begin{equation}\label{17a}
f^\alpha = i_v (F^\alpha - * F_\beta \epsilon^{\alpha\beta}),
\end{equation}
where $i_v$ denotes the contraction with the vector field correponding to $v$, 
one can verify that the action
\begin{eqnarray}\nonumber
S_{PST}(A^\alpha, C^\alpha)={1\over 4}&\int& (F^\alpha *F^\alpha + f^\alpha * 
f^\alpha) \\
+ {1\over 2}&\int&(C^1 dA^2 - dA^1 C^2)\label{18}
\end{eqnarray}
gives the duality condition as equation of motion, besides the "Maxwell" 
equations.
The analogous action  for the self-dual case $F= * F$ is obtained 
identifying $A^1 = A^2$, $C^1 = C^2$ and it is given by
\begin{equation}
S_{PST} = {1\over 2} \int (F *F + f* f) + \int C d A.
\end{equation}

The basic reason why the PST formalism works correctly is the existence of 
a (PST)-symmetry 
\begin{equation}
\delta A^\alpha = - {\varphi \over ||da||} f^\alpha, \quad \delta a= \varphi,
\end{equation}
allowing to choose the 0-form $a$ arbitrarily, so that it does not propagate 
unphysical degrees of freedom, in addition to a symmetry 
\begin{equation}
\delta A^\alpha = \Phi^\alpha da
\end{equation}
allowing to reduce the second order equations of motions for the gauge 
fields to the first order duality equation.  $\varphi$ and $\Phi^\alpha$ 
are transformation parameters.  
Together with the invariance under  Dirac brane changes, the PST symmetries 
fix the action completely in the self-dual case, leaving the freedom to add 
a term
\begin{equation}
{1\over 2} \int \ C^1 C^2
\end{equation}
in the dual-case.

Furthermore if the invariant curvatures have the more general structure 
\begin{equation}\label{25}
F^\alpha = dA^\alpha + C^\alpha + L^\alpha,
\end{equation}
where $L^\alpha$ are fields not transforming under Dirac brane changes, then 
the above symmetries identify the invariant action as 
\begin{equation}
S_{PST}(A^\alpha, C^\alpha + L^\alpha) + {1\over 2} \int (C^1 L^2 - L^1 C^2).
\end{equation} 

Finally let us make a remark on the nature of the coupling $C$-$A$ in PST. 

To get an intuitive idea consider the case where $C_2 = j_2= 0$ then, 
a "magnetic coupling" $C$-$A$ in the action would have the structure 

\begin{equation}
{1\over 2} \int F* F= {1\over 2} \int (dA + C) * (dA + C)
\end{equation}
whereas  on "electric coupling" would be 
\begin{equation}
\int d A C = \int A d C = \int A j.
\end{equation}

In the PST approach instead we have
\begin{equation}
{1\over 4} \int F* F + {1\over 2} \int d AC
\end{equation}
hence it is a kind of "dyonic" $C$-$A$ coupling with "${1\over 2}$ 
magnetic coupling" 
and " ${1\over 2}$ electric coupling".

\section{Dirac branes approach to p-brane systems}
To every (elementary) closed $p$-brane we associate a $d-(p+1)$ current 
$j_{d-p+1}$ describing the worldvolume $\Sigma_{p+1}$ of the brane, via 
$PD$: $j_{d-p+1}= PD 
(\Sigma_{p+1})$.
Since $j \equiv j_{d-p+1}$ is closed we can associated a $(p+1)$ - Dirac 
brane via: 
\begin{equation}
\Sigma_{p+1} = 
\partial \Sigma_{p+2}, PD (\Sigma_{p+2}) = C_{d-(p+2)} \equiv C,
\end{equation}
so that $j=dC$.

The $p$-brane interacts with a gauge $(p+1)$-form either electrically, and 
then no Dirac brane is needed, or magnetically, and then in our formalism 
we have an associated Dirac brane.
In particular if a $p$-brane is electrically coupled, its dual 
$(d-(p+4))$-brane is magnetically coupled.
In the PST formalism, due to the "dyonic coupling", Dirac branes are 
always needed.

One introduces a manifestly Lorentz-invariant action describing the 
interaction between gauge forms, $p$-branes and $(p+1)$-Dirac branes via 
the PST formalism (or the related Schwinger \cite{Schw} or Zwanziger  
\cite{Zwan} formalisms, see \cite{Kurt}). This procedure requires as 
consistency condition the indipendence of the choice of the Dirac branes. 
This put strong restrictions, such as Dirac quantization conditions, which 
in turn might generate new physics like spin-statistics transmutation for 
dyons in $d=4$ \cite{Ian} .
(This transmutation has been discussed at quantum field theory level in 
\cite{Marc}, combining a formalism involving Mandelstam strings, developed 
in \cite{Froh}, with 
Schwinger or PST techniques). 

\section{Anomaly cancellation in 5-branes}

Let us apply the formalism sketched above to analyze the anomaly 
cancellations in 5-branes. 

We discuss in particular the system SUGRA 10+ heterotic 5-brane + heterotic 
string \cite{Witt}. 
The gauge $p$-forms of SUGRA are denoted by $B_2$ and $B_6$, the current 
associated to the 5-brane by $J_4= PD (\Sigma_6$) and its Dirac brane 
by $C_3$ and the current associated to the heterotic string 
by $J_8 =PD (\Sigma_2)$ and its Dirac brane by $C_7$.
We denote the $SO(1,9)$ curvature by $R$, the $SO(32) \bigotimes SU (2)$ 
curvature by $F \bigoplus G$ and the curvature of the normal $SO(4)$ 
bundle $N$ of the 5-brane by $T$. In terms of these curvatures one defines
\begin{eqnarray}
&X_8& = {1\over 192 (2\pi)^4} (tr R^4 + {1\over 4} (tr R^2)^2 \nonumber\\
 &-&  tr  R^2 tr F^2 + 8 tr F^4) \\
&X_4 & ={1\over 4(2\pi)^2} (tr R^2 - tr F^2) \\
&Y_4 & = {1\over 48 (2\pi)^2} (tr R^2 - 2 tr T^2 - 24 tr G^2) \\
&\chi_4& = {1\over 8(2\pi)^2} \epsilon^{a_1 a_2 a_3 a_4} T^{a_1 a_2} 
T^{a_3 a_4}, 
\label{28}
\end{eqnarray}
where $\chi_4$ is the Euler form of $N$.
The invariant polynomials are then given by $X_8 X_4$ (SUGRA), 
$X_8+ (X_4+\chi_4) Y_4$ (5-brane), $X_4$ (string), a factor $2\pi$ being 
omitted. With the standard convention, e.g.
\begin{equation}
X_8 = d X_7, \quad \delta X_7 = d X_6^1,
\end{equation} 
where $\delta$ denotes the gauge variation, the anomaly polynomials 
obtained via transgression are given by $X^1_6 X_4$ (SUGRA), 
$X_6^1 + (X^1_2+\chi^1_2) Y_4 J_4$ (5-brane) $X^1_2 J_8$ (string).

The cancellation of anomalies is due to Green-Schwarz \cite{Green} plus
inflow \cite{Call} mechanisms. 
In the standard treatment the cancelling terms in the lagrangian are 
$B_2 X_8$ (SUGRA), $(B_6+B_2 Y_4) J_4$ (5-brane), $B_2 J_8$ (string) with the 
gauge transformations: 
\begin{equation}
\delta B_2 = -X^1_2, \quad \delta B_6 = X^1_6,
\end{equation}
but with an additional contribution on the 5-brane given by 
\begin{equation}\label{36}
(\delta B_2)|_{\Sigma_6}= -\chi^1_2
\end{equation}
to cancel $\chi^1_2 Y_4 J_4$; this last transformation corresponds to the 
inflow.

However a clear-cut definition strictly localized on the 5-brane appears 
somewhat cumbersome, in fact if the additional transformation (\ref{36}) 
is added with a characteristic function of $\Sigma_6$, then from the point 
of view of distribution theory it is irrelevant,
because $\Sigma_6$ is of measure zero in $\textbf{R}^{10}$ and the 
characteristic function is bounded, on the other hand if it is added with 
a $\delta$-function on $\Sigma_6$, then it modifies the equations of motion.
A solution is obtained via a smoothing operation, mathematically using 
Thom forms
\cite{bott}, but then this contribution is no more localized strictly on the 
5-brane.

The Dirac brane approach automatically solves this problem \cite{Lech}. In
fact, we 
start from the equations for the invariant curvatures

\begin{eqnarray}
& dH_3&  = X_4 + J_4 \\  
& dH_7& = X_8 + Y_4 J_4 + J_8 \\ 
& H_3 & = *H_7
\end{eqnarray}
and we
write $X_4 = d X_3, X_8= dX_7$, $Y_4 =d Y_3$, where $X_3, X_7, Y_3$ are 
the corresponding  Chern-Simons forms, and 
\begin{equation}
h_2 = i_v (H_3 - *H_7), \quad h_6 = i_v (H_7 - *H_3).
\end{equation} 
According to the rules outlined in previous section we pose
\begin{eqnarray}
&H_3& = dB_2 + X_3 + C_3 \\
&H_7& = dB_6+ X_7 + Y_3 J_4 + C_7,\nonumber
\end{eqnarray}
where $X_3$ and $X_7 + Y_3 J_4$ play the role of $L^\alpha$ in the 
equation (\ref{25}). 
In quadratic approximation for $h_2$ and $h_6$, but 
a Born-Infeld action can be
easily arranged \cite{Soro}, 
the relevant part of the lagrangian of the coupled system  is given by:
\begin{eqnarray}\nonumber
&{1\over 4}& (H_3 *H_3 + H_7 *H_7 + h_2 *h_2 + h_6 *h_6)  \\
& + {1\over 2}& (X_7 + Y_3 J_4 + C_7) d B_2 + {1\over 2} (X_3 + C_3) 
dB_6 \nonumber \\
& + {1\over 2}& (X_7 + Y_3 J_4) C_3 + {1\over 2}X_3 C_7 + 
{1\over 2} C_3 C_7. \label{}
\end{eqnarray}

The gauge transformations are given by

\begin{equation}
\delta B_2 = - X^1_2, \quad \delta B_6 = - X^1_6 - Y^1_2 J_4,
\end{equation}
completely free of ambiguities.
The corresponding contribution to the anomaly is given by, 

\begin{eqnarray}\nonumber
&-& \int [{1\over 2} ( X^1_2 X_8 + X^1_6 X_4)] 
+ X^1_2 J_8 \\
&+& (X^1_6 + {1\over 2} (X^1_2 Y_4 + Y^1_2 X_4) + Y^1_2 J_4) J_4 \nonumber\\
\end{eqnarray}
which is equal, up to trivial cocycles, to

\begin{equation}\label{46}
-\int [X^1_6  X_4 + X^1_2 J_8+ (X^1_6 + X^1_2 Y_4 + Y_2^1 J_4)J_4].
\end{equation}
The last term in (\ref{46}) is naively ill-defined, but using the theory of 
characteristic currents developed above (i.e. replacing $J_4 J_4$ 
by $\lim_{\epsilon\rightarrow 0}
J^\epsilon_4 J^\epsilon_4$) gives the desired term $Y^1_2 \chi_4 J_4 
= \chi^1_2 Y_4 J_4$+ trivial cocycle.
Hence we obtain the cancellation of anomalies with the inflow contribution 
strictly localized on the 5-brane.

\section{Some remarks on M5}
An action for the coupled system SUGRA 11 + M5-brane was first proposed 
in \cite{Bando}, but it fails to exhibit Dirac brane independence and does 
not consider the anomaly cancellation.
Anomaly cancellation was achieved in \cite{Freed} but via smoothing of 
the M5-brane worldvolume, using a regular Thom form, and this might cause 
some problem if one considers the coupling also to the M2-brane, due to a 
violation of Dirac quantization for fluxes, introduced by the non-locality 
involved in the construction. A local mechanism for anomaly cancellation 
was proposed in \cite{Bono}, but it needs reducible connections in 
the $SO(5)$-normal bundle of the 5-brane and the complete action for the 
coupled system is not discussed;  a residual dependence of a 5-vector field on
M5 remains in this approach.  

We reconsider this problem in collaboration with M. Tonin \cite{Toni} using 
the formalism of Dirac branes and characteristic currents. We analyze the 
complete coupled system and we achieve a complete cancellation of gauge and 
gravitational anomalies, in local form. 
However, as in \cite{Bono}, it remains a dependence of the direction along 
which the  worldvolume of the Dirac brane, $\Sigma_7$, is attached to the 
worldvolume, $\Sigma_6=\partial \Sigma_7$, of the M5-brane.
This dependence however, completely disappears in the system of NS-5-brane 
coupled to the II A SUGRA 10, treated with our formalism, a priori 
corresponding to a dimensional reduction.
For details of our analysis of the M5-brane see \cite{Toni}, here we 
simply sketch how the formalism of characteristic currents helps in the game.
One of the new features appearing in the treatment of this system is the 
Chern-Simons 
term $A_3 dA_3 dA_3$, where $A_3$ is the gauge 3-form  of SUGRA 11.  When 
a coupling 
with an M5-brane is considered, it is clear how to modify $dA_3$: simply 
adding the Poincar\'e dual of the worldvolume associated to the 6-Dirac 
brane, $C_4 = PD (\Sigma_7)$, but it is less clear how to modify 
$A_3|_{\Sigma_7}$  since $C_4$ is not a closed form. However, using the 
theory of characteristic currents, replacing $C_4$ by $C_4^\epsilon$, 
we have a natural candidate, because the pull-back of $C_4^\epsilon$ 
to $\Sigma_7$ is given by $\chi_4$, the Euler form of the normal bundle of 
the Dirac 
6-brane. $\chi_4$ is a closed form, hence locally $\chi_4= d\chi_3$, 
and $\chi_3$ is the natural candidate we are looking for. 

Finally we remark a suggestive feature of the formalism: a priori it 
is consistent with the modified Dirac quantization condition \cite{Phys} 
for the invariant curvature $H_4$ $= dA_3 + C_4$. In fact $H_4|_{\Sigma_6}$ 
involves a term $C_4 |_{\Sigma_6}$, naively ill-defined.
Using the characteristic currents we replace it by 
\begin{equation}
\lim_{\epsilon\rightarrow 0} C_4^\epsilon |_{\Sigma_6} = 
{\chi_4\over 2} |_{\Sigma_6}
\end{equation} 
where $\chi_4$ is the Euler characteristic of the $SO(4)$-bundle of the 
Dirac brane. Hence for a 4-cycle $Z_4$ in $\Sigma_6$ we have 
\begin{equation}  
\int_{Z_4} H_4 = \int_{Z_4} {\chi_4\over 2} \in {\textbf{Z}\over 2},   
\end{equation}
so that the cohomology class of $H_4$ is  a priori half-integer valued, 
as discussed in \cite{Phys}.

{\bf Acknowledgments.}
We thank M.Tonin for the joy of collaboration and I. Bandos, P. Pasti 
and D. Sorokin
for useful discussions.

\end{document}